# Monolithic low-noise erbium-doped thin-film lithium niobate waveguide amplifier with 18 dB fiber to fiber net gain

Mengqi Li[1,2], Jianping Yu[2], Zhe Wang[2], Botao Fu[2], Rongbo Wu[2], Min Wang[2], Haisu Zhang[2,*], and Ya Cheng[1,2,3,4,5,6,*]

[1]State Key Laboratory of Precision Spectroscopy, East China Normal University, Shanghai 200062, China.
[2]The Extreme Optoelectromechanics Laboratory (XXL), School of Physics and Electronic Sciences, East China Normal University, Shanghai 200241, China.
[3]Hefei National Laboratory, Hefei 230088, China.
[4]Shanghai Research Center for Quantum Sciences, Shanghai 201315, China.
[5]Collaborative Innovation Center of Extreme Optics, Shanxi University, Taiyuan 030006, China.
[6]Collaborative Innovation Center of Light Manipulations and Applications, Shandong Normal University, Jinan 250358, China.
*Correspondence: H. Zhang (hszhang@phy.ecnu.edu.cn), Y. Cheng (ya.cheng@siom.ac.cn).

## Abstract

Erbium-doped waveguide amplifiers have captured great attentions in recent years due to the rapid advance of photonic integration materials and fabrication techniques. In this work, a compact erbium-doped thin-film lithium niobate waveguide amplifier integrated with high-efficiency edge-couplers on the small footprint of 2 mm×25 mm, achieving >18 dB fiber-to-fiber (off-chip) net gain with bidirectional pumping by ~1480 nm laser diodes, is fabricated by the photolithography assisted chemo-mechanical etching technique. The fiber-to-fiber noise figures of the amplifier are also characterized to be around 5 dB, and the maximum amplified signal powers at the output fiber are above 13 dBm. Theoretical amplifier modeling resolving the erbium absorption and emission spectra predicts the efficient gain scaling with waveguide length for most of the telecom C-band wavelengths. The demonstrated high-external-gain erbium-doped waveguide amplifier will benefit various applications from optical communication and metrology to integrated photonic computing and artificial intelligence.

## Introduction

Erbium-doped waveguide amplifiers and lasers have been broadly investigated in recent decades [1-4]. Due to the stable optical transitions and long excited-state lifetimes of rare-earth-ions (REIs) doped in suitable host materials, broadband light emissions and amplifications are reliably achieved [5]. Benefited from the spatially and temporally stable optical gain of erbium ions ($Er^{3+}$) and the confined optical guiding structures, erbium-doped waveguides are ideal platforms for optical amplifiers and lasers in the important telecom C-band, with the renowned example of erbium-doped fiber amplifiers (EDFAs) widely used in optical communication and metrology [6,7]. The recent advance of photonic integration circuits (PICs) stimulates the desire for miniaturized erbium-doped waveguide amplifiers (EDWAs) monolithically integrated on a chip, which feature high-scalability and low-power consumption as well as the inherent low-crosstalk multi-wavelength amplification facilitated by erbium ions [8-10]. Various material platforms have been adopted for PIC-based EDWAs with the most frequently studied ones as alumina oxide ($Al_2O_3$), silicon nitride ($Si_3N_4$), and thin-film lithium niobate (TFLN) [11-25]. The former two material platforms hold the advantage of matured CMOS-compatible fabrication procedures while the latter one which emerges in the recent decade still lacks standard batch production techniques.

Owing to the excellent optical and nonlinear optical properties of the ferroelectric lithium niobate crystals, TFLN-based PIC researches have been boomed in recent years with main focus on its high-speed electro-optical tuning ability in light modulators, beam scanners and optical computing [26, 27]. On the other hand, lithium niobate crystals are also ideal host materials for rare-earth-ions due to their intrinsic lattice defects [28]. The advent of REI-doped TFLN stimulates broad interests for on-chip integrated micro-lasers, optical amplifiers and quantum memories [29]. Among the plethora of demonstrations TFLN-based EDWAs have been continuously studied aiming to achieve comparable gain and output powers with standard EDFAs. Various TFLN-EDWAs have been proved with high on-chip gains and TFLN-based power amplifiers utilizing large-mode-area (LMA) waveguides are also tested [17-25]. Nevertheless, due to the inherent nonlinearity of the single-crystalline lithium niobate and the obstacles of efficient light coupling into TFLN waveguides, the directly obtained off-chip gains and powers are rather limited in the established TFLN-EDWAs, impeding their practical application in fiber communication networks. A recent work adopting LMA-waveguides and high-efficiency edge-couplers achieves the off-chip net gain of 10 dB, though the employed straight waveguide structure limits the usable length and noise figure at the expense of large chip size [30].

In this work, a compact TFLN-EDWA with the small footprint of 2 mm×25 mm and the total waveguide length of 8.5 cm is designed and fabricated by the photolithography assisted chemo-mechanical etching (PLACE) technique. Segmented LMA-waveguides are connected with adiabatic waveguide tapers and single-mode waveguide bends. Specifically, the EDWA-chip are integrated with index-matched edge-couplers which enables high-efficiency coupling with external optical fibers. The fiber-to-fiber (off-chip) net gain of the fabricated amplifier is systematically investigated, giving the largest value of 18.1 dB at the signal wavelength around 1532 nm and the fiber input power of -20 dBm. Meanwhile, the fiber-to-fiber noise figures are characterized by the optical source subtraction method to be around 5 dB. To the best of our knowledge, this is the first demonstration of low-noise high-external-gain TFLN-EDWA. Besides, the amplifier gain saturation is measured giving the maximum amplified signal power in the output fiber to be above 13 dBm. Furthermore, the broadband amplification property of the TFLN-EDWA is tested as well. Theoretical amplifier modeling considering the wavelength-dependent erbium absorption and emission is conducted to reveal the gain-scaling with waveguide length at the longer wavelengths of the telecom

C-band. The demonstrated low-noise high-external-gain TFLN-EDWA with compact footprint holds great promise in various applications such as optical coherent communication and integrated laser.

## Results

Design and fabrication of the TFLN-EDWA

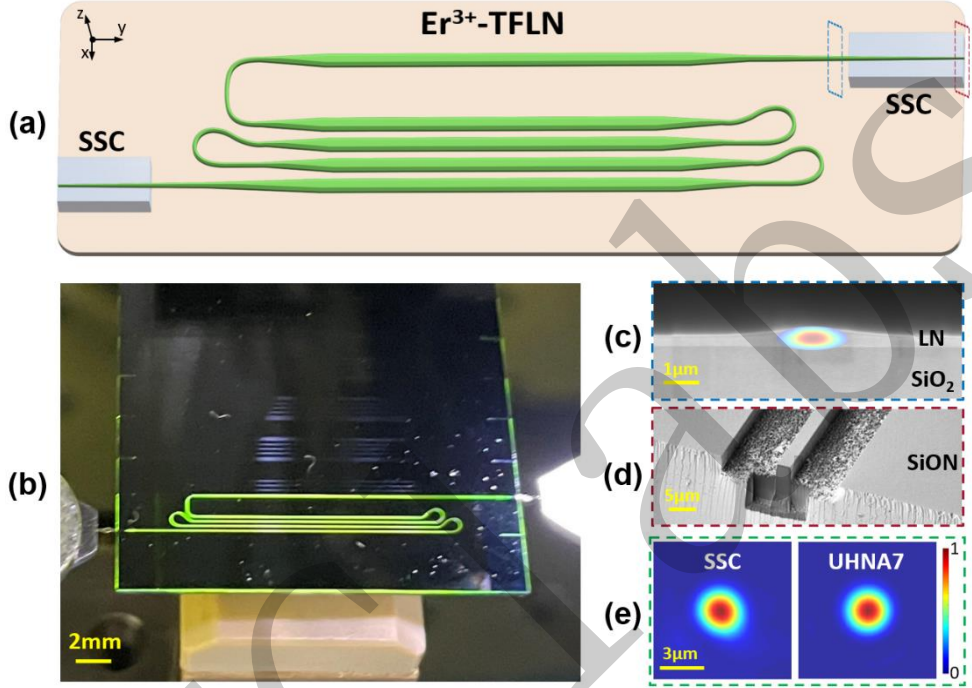

Fig. 1. TFLN-EDWA chip design. (a) Schematic of the EDWA chip composed of TFLN waveguides and SSC edge-couplers. (b) Digital camera image of the EDWA chip butt-coupled with optical fibers. (c) SEM cross-section image of the TFLN waveguide. (d) SEM cross-section image of the SSC-port. (e) Infrared microscope images of the output mode profiles for the SSC-port and the UHNA7 fiber.

The TFLN-EDWA is fabricated on the erbium-doped Z-cut lithium niobate on insulator (LNOI) wafer, which is prepared by the ion-slicing process (NanoLN, Jinan Jingzheng) from the congruent erbium-doped lithium niobate bulk crystal grown by the Czochralski method. The erbium-doping concentration is about 0.5 mol%. The LNOI wafer consists of 500-nm-thick top LN layer, 4.7-μm-thick buried oxide layer and 500-μm-thick bottom silicon substrate. The TFLN waveguide is fabricated by the PLACE technique consisting of femtosecond laser patterning and chemo-mechanical polishing [31]. The designed footprint of the EDWA chip is shown in Fig. 1(a). Waveguide bends with minimum radius of 400 μm are employed to increase the waveguide length at the small chip size of 2 mm×25 mm. The total waveguide length is around 8.5 cm. In the central region of the chip straight LMA-waveguides with the top width of 10 μm are employed to reduce the nonlinear impairment at high input powers. The LMA-waveguides are adiabatically tapered to the narrow waveguides of 1 μm top width before the bend regions. The waveguide bend profile is optimized using the combination of circular-bend and Euler-bend [32]. Moreover, the spot-size-convertor (SSC) based edge-couplers are integrated at both waveguide ends for high-efficiency coupling with external optical fibers. The SSC structure enables the robust mode conversion from the underneath TFLN waveguide to the overcladding SiON waveguide. Detailed information about the SSC edge-couplers can be found in our previous works [30, 33].

The fabricated EDWA chip is shown in Fig. 1(b). Bright up-conversion fluorescence from erbium ions excited by the pump laser is clearly observed along the full waveguide path. The scanning-electron-microscope (SEM) images for the cross-section of the TFLN waveguide (at the blue dashed square in Fig. 1(a)) is shown in Fig. 1(c), where a ridge profile with slant sidewalls typical of the PLACE fabrication technique can be clearly seen. The simulated intensity distribution for the fundamental transverse-electric mode ($TE_{00}$) is superposed on the waveguide cross-section image in Fig. 1(c). Furthermore, the SEM cross-section image for the SSC output port (annotated by the red dashed square in Fig. 1(a)) is shown in Fig. 1(d). The SSC-port is prepared by focused-ion-beam (FIB) milling for better clarity. The cross-section scales of the SSC-port feature the 3 μm×3 μm square shape of SiON waveguide, whose output mode profile at the wavelength of 1550 nm is measured and shown in the left panel of Fig. 1(e). To maximize the coupling rate from optical fiber to SSC-port, the ultrahigh-NA fiber (UHNA7) which has the NA=0.41 and a mode field diameter (MFD) of 3.2 μm at 1550 nm is employed for butt-coupling with the chip. The recorded output mode profile of the UHNA7 fiber is also depicted in the right panel of Fig. 1(e). The modal overlap between the mode profiles of the SSC-port and the UHNA7 fiber is estimated to be >90%.

## Experimental characterization of the EDWA chip

The experimental setup for characterization of the TFLN-EDWA chip is shown in Fig. 2(a). Two fiber-coupled high-power laser diodes at the wavelength of 1480 nm are used as the pumps, and a C-band tunable external cavity continuous-wave diode laser (Toptica) is used as the signal. The signal laser and one pump laser are combined by the fiber-based wavelength division multiplexer (WDM) and then connected to the standard single-mode fiber (SMF) which is fusion-spliced with the UHNA7 fiber. The splicing loss between the UHNA7 fiber and the SMF fiber is around 1 dB. The UHNA7 fiber is positioned in close contact with the SSC-port of the EDWA chip by a high-resolution three-dimensional motorized stage, and index-matching gels are used at the fiber-chip interface for Fresnel-reflection suppression. The facet coupling loss between the UHNA7 fiber and the SSC-port is measured to be less than 2 dB at 1550 nm. Thus, the lumped coupling loss from the SMF to the EDWA chip is around 3 dB/facet. Similar with the first light path the other pump laser is routed through the fiber-WDM and the SMF-spliced UHNA7 fiber, and then injected into the EDWA chip. The output signal is measured after the second fiber-WDM by the optical spectrum analyzer (OSA), which is pre-calibrated with high-resolution power meters. Fiber-based polarization controllers (PC) are inserted in both pump laser and signal laser paths for excitation of the transverse-electric (TE) modes in the TFLN waveguides.

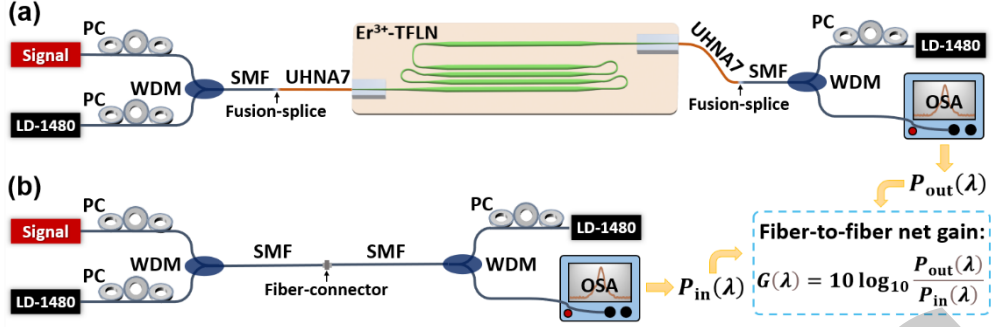

Fig. 2. Experimental setup. (a) Schematic of the setup for characterizing the fiber-to-fiber net gain of the TFLN-EDWA chip. PC: fiber-based polarization controller, WDM: fiber-based wavelength division multiplexer, OSA: optical spectrum analyzer. (b) Schematic of the setup for measuring the input signal power. The definition of fiber-to-fiber net gain is shown in the inset.

In order to retrieve the fiber-to-fiber off-chip gain of the EDWA chip with high accuracy, the input and output signal powers are measured using the same fiber connection and equipment to avoid uncontrolled experimental uncertainty. As shown in Fig. 2(b), the input signal power in the SMF is measured by directly connecting the input SMF with the output SMF, passing the second WDM and then recorded by the OSA as $P_{\text{in}}(\lambda)$. Similarly, the EDWA chip's amplified signal power in the output SMF is also measured by passing the second WDM and recorded by the OSA as $P_{\text{out}}(\lambda)$. The fiber-to-fiber net gain of the EDWA chip is defined as:

$$G(\lambda) = 10 \log_{10} \frac{P_{\text{out}}(\lambda)}{P_{\text{in}}(\lambda)} = G_0(\lambda) - 2\alpha_{\text{sl}}(\lambda) - 2\alpha_{\text{c}}(\lambda) \qquad (1)$$

By the procedure described above the off-chip gain $G(\lambda)$ is actually defined as the external gain from input SMF to the output SMF, and the usually used on-chip gain $G_0(\lambda)$ can be deduced by compensating the lumped coupling loss from SMF to the EDWA chip (including the splicing loss $\alpha_{\text{sl}}(\lambda)$ between SMF and UHNA7 and the coupling loss $\alpha_{\text{c}}(\lambda)$ between UHNA7 and the SSC-port), as also shown in the right part of the above equation.

The fiber-to-fiber net gain of the EDWA chip is first measured at the signal wavelength around 1532 nm and the input power of -20 dBm. The output signal spectrum is shown in red color in Fig. 3(a). The input pump power is 200 mW. To visualize the off-chip gain the input signal spectrum is also shown in green color in Fig. 3(a). For comparison the output signal spectrum without pump is depicted in blue color as well. It can be clearly seen that upon amplification by the EDWA chip, a broad background noise arising from the amplified spontaneous emission (ASE) of erbium ions accompanies the amplified signal. The ASE-noise can be filtered by narrow-linewidth bandpass filters at the signal wavelength, though the residue noise within the signal bandwidth is not easy to remove and its relative strength defines the amplifier's noise figure as described later. The spectra in Fig. 3(a) are enlarged around the signal wavelength and shown in Fig. 3(b). It can be clearly seen that the difference between the input and output signal power-levels indicates the net gain G=18 dB, while the usually defined signal enhancement (SE) factor by comparing the output signal powers with and without the pump is SE=78 dB. Furthermore, the amplifier fiber-to-fiber noise figures (NF) are deduced using optical source subtraction method as:

$$\text{NF} = \frac{P_{\text{so}} - GP_{\text{sse}}}{Gh\nu B_0} + \frac{1}{G} \qquad (2)$$

where $P_{\text{sse}}$ and $P_{\text{so}}$ are the input and output noise power within the optical bandwidth of $B_0$ at the signal frequency $\nu$, $G$ is the net gain, and $h$ is the Planck constant [16]. The retrieved noise powers from input and output signal spectra are $P_{\text{sse}}$=-75 dBm and $P_{\text{so}}$=-34.3 dBm, and the resolution bandwidth of OSA is 0.116 nm which gives $B_0$=14.8 GHz. Substituting the above parameters into Eq. (2) the noise figure in this case is calculated as NF=4.9 dB.

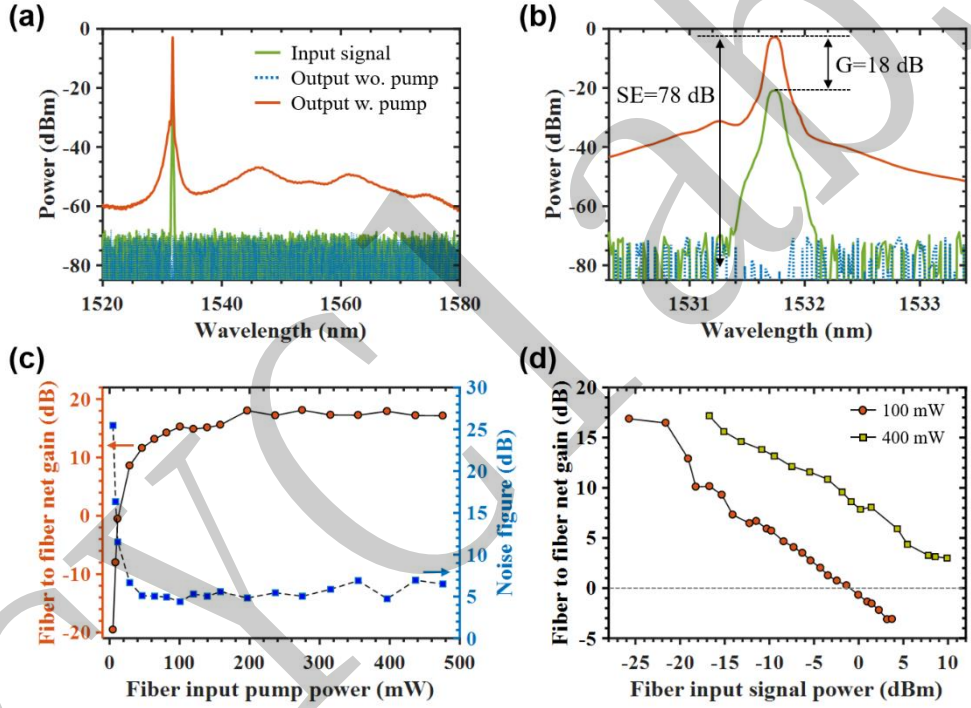

Fig. 3. Experimental results for single-wavelength amplification. (a) The input and output signal spectra of the EDWA chip. (b) Enlarged view of the curves in (a) with the annotations of net gain and signal enhancement. (c) The fiber-to-fiber net gains and noise figures with respect to input pump powers. (d) The fiber-to-fiber net gains at variable input signal powers. The grey dashed line denotes the transparency level for the fiber-chip-fiber link.

The fiber-to-fiber net gain and noise figure of the EDWA chip is further measured at variable pump powers and the results are shown in Fig. 3(c). The off-chip net gain emerges when the input pump power is larger than 10 mW and increases rapidly until the input pump power of 200 mW, and then the gain saturates around 18 dB with further pump power increments. In the meantime, the amplifier noise figures decrease with the increasing pump powers below 300 mW, from NF=17 dB at G=0 dB to around 5 dB at the gain saturation region. Further pump power increment induces higher noise figures though the amplifier gain is almost constant. So, for the high-gain small-signal amplification the pump power should be kept at modest levels to avoid the additional amplifier noise.

The signal gain saturation properties of the EDWA chip are further characterized at increasing signal powers and constant pump powers. Two pump power-levels are selected for the gain saturation measurements. The results are shown in Fig. 3(d). At the pump power of 100 mW (red circles in Fig. 3(d)), the fiber-to-fiber net gain decreases quickly when the input signal power increases from -25 dBm to 5 dBm, and the zero-gain point corresponds to the input signal power around 0 dBm. To obtain higher gain saturation level, the pump power is elevated to 400 mW and the signal power is increased from -17 dBm to 10 dBm (yellow squares in Fig. 3(d)), the gain decays slower in this case and the gain for the input signal power of 0 dBm is about 8 dB. The maximum amplified signal output power is about 13 dBm with the input signal power of 10 dBm.

## Broadband light amplification

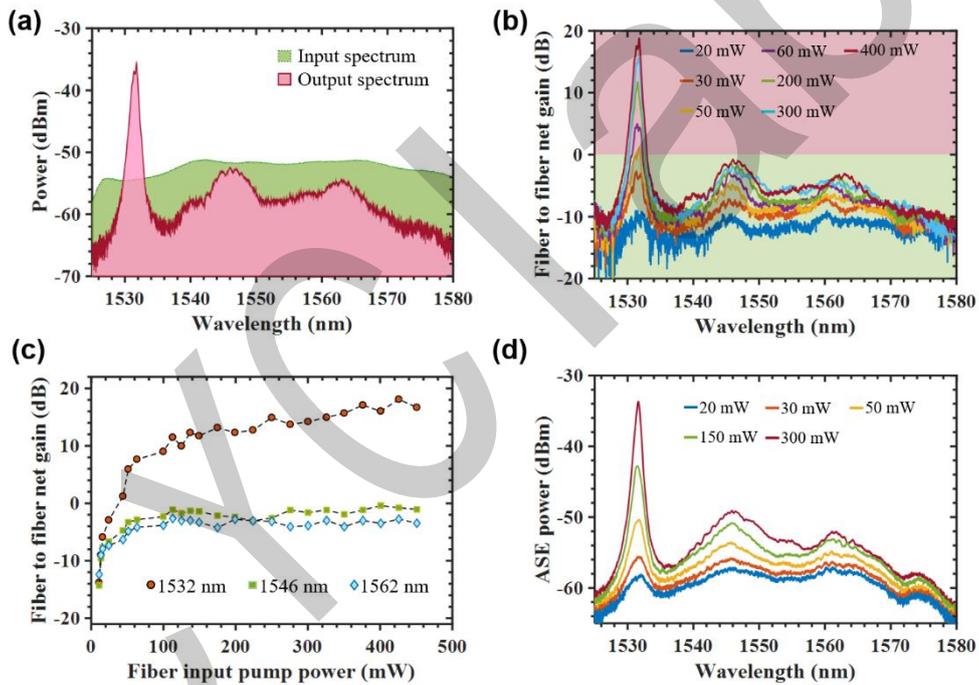

Fig. 4. Experimental results for multi-wavelength amplification. (a) The input and output signal spectra of the EDWA chip. (b) The gain spectra variations with respect to the input pump powers. (c) The fiber-to-fiber gains at the selected wavelengths with respect to the input pump powers. (d) The ASE spectra of the EDWA chip at variable input pump powers.

To test the multi-wavelength amplification property of the EDWA chip, a broadband light source from the amplified spontaneous emission of the commercial EDFA (KY-ASE-CL-10, Beijing Keyang) is employed as the input signal. To avoid gain competition between different wavelengths the input signal power is reduced to -20 dBm. The measured spectrum for the input signal is shown in Fig. 4(a) as the green shaded curve, and a broadband spectrum spanning from 1525 nm to 1580 nm can be noticed. The typical output spectrum from the EDWA chip is further shown as the red shaded curve in Fig. 4(a). It can be clearly seen that the output spectral components around 1532 nm are significantly higher than the input components, which means that the fiber-to-fiber net gain is maximum in this wavelength region. The wavelength dependent net gain can be obtained by subtracting the output spectrum with the input spectrum, and the deduced gain spectra at variable pump powers are shown in Fig. 4(b).

It can be clearly seen from Fig. 4(b) that the net gain at 1532 nm grows faster following increased pump powers, and the other components within the input signal spectra increase much less with pump powers. The gain variations at the wavelengths of 1532 nm, 1546 nm and 1562 nm are further plotted in Fig. 4(c). These wavelengths are selected in accordance with the three prominent fluorescence peaks of erbium ions. Due to the maximum stimulated emission cross section at 1532 nm, the gain at this wavelength will increase quickly as soon as the sufficient population inversion is realized at high pump powers, which is well-revealed by the red squares in Fig. 4(c). On the contrary, the gain at the other two wavelengths saturates at much lower values due to the inadequate emission probabilities of erbium ions at these wavelengths in spite of large population inversion at high pump powers. To increase the amplifier gain outside the 1532 nm region, higher erbium concentrations and longer waveguide lengths are needed.

Besides the amplification of the broadband signal, the ASE-emissions from the EDWA chip are also recorded and shown in Fig. 4(d). For ASE measurements only pump lasers are injected into the chip. The ASE spectra evolution in Fig. 4(d) is similar to the gain spectra variations shown in Fig. 4(b). Since both the gain spectra and the ASE spectra depend on the average population inversion of the EDWA chip, the consistent results from both measurements indicate that the inversion-fractions increase with pump powers and a large inversion-fraction with suppressed ground-state absorption is achieved at high pump powers.

## Theoretical modeling and discussions

Due to the dominating homogeneous broadening of $Er^{3+}$ in $LiNbO_3$ crystals (the inhomogeneous broadening is only 180 GHz while the homogeneous broadening is above 500 GHz at room temperature) [34, 35], the different Stark-level transitions between the excited-state $^4I_{13/2}$ and the ground-state $^4I_{15/2}$ are less overlapped, compared to the case of $Er^{3+}$ in amorphous and polycrystalline host materials like silica and $Al_2O_3$. As a result, the absorption and emission spectra of $Er^{3+}$:$LiNbO_3$ feature the intense peak around the wavelength of 1532 nm and the weak multi-shoulder pedestal corresponding to resonant transitions between different Stark-levels. This spiky spectrum renders a few characteristics for the waveguide amplifier based on $Er^{3+}$:$LiNbO_3$. First of all, the peak gain at 1532 nm will dominate the gain spectrum at sufficient pump powers, and the demonstrated TFLN-EDWAs mostly fall in this scenario [17-25]. To increase the overall gain at broad wavelength range, longer waveguide lengths have to be used.

Caution should be paid here that the gain at 1532 nm will always grows the fastest within the gain spectrum, and to some extent the back-reflected light at this wavelength by material defects and waveguide ends will induce parasitic-lasing effect, which can in turn deplete the excited-state population and reduce the attainable gain for other wavelengths [36, 37]. Thus, the return-loss (the loss of reflected light power) should be increased at long waveguide lengths. Besides, the gain equalization during multi-wavelength amplification will be more challenging for $Er^{3+}$:$LiNbO_3$ based amplifiers due to the distinct absorption and emission cross sections at each wavelength channel and the homogeneous saturation effect [8].

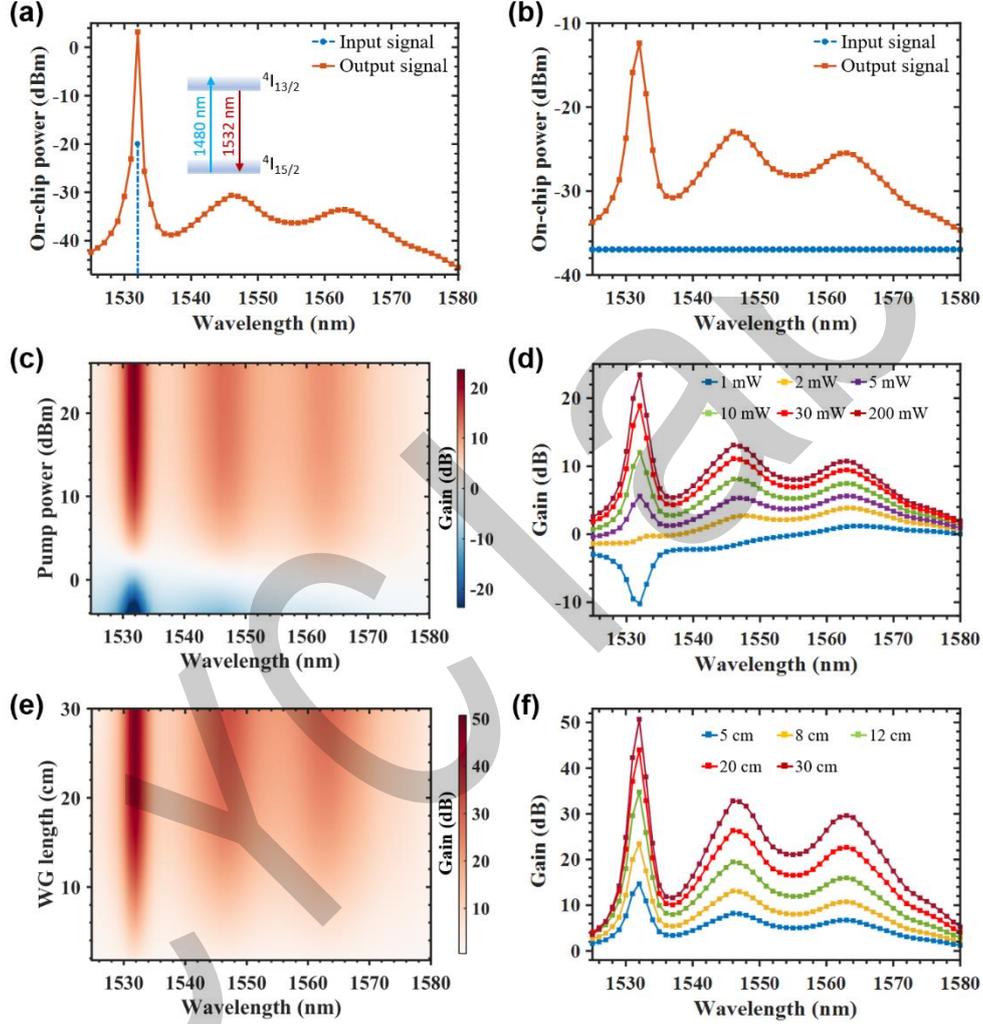

Fig. 5. Simulation results. (a) The input and output spectra for single-wavelength signal amplification. The energy-levels of $Er^{3+}$ used in the simulation is shown in the inset. (b) The input and output spectra for multi-wavelength signal amplification. (c) The pseudo-color plots of the gain spectra evolution with respect to input pump powers. (d) The gain spectra at different pump powers. (e) The pseudo-color plots of the gain spectra evolution with respect to EDWA waveguide lengths. (f) The gain spectra at different waveguide lengths. The waveguide length is 8 cm for (c) and (d), and the pump power is 200 mW for (e) and (f). Spectral resolution of 1 nm is used in the simulation.

To get a better understanding of the impact of $Er^{3+}$:$LiNbO_3$ spectra on the EDWA's performance, the amplifier model involving the static rate equations and the wavelength-dependent absorption and emission spectra is employed [38, 39]. The population dynamics under the two-level approximation: $^4I_{13/2}$ and $^4I_{15/2}$ (see the inset in Fig. 5(a)), is used to describe the absorption and stimulated emission of $Er^{3+}$ induced by the pump and signal lights. The background ASE noise is also included by adding the spontaneous emission terms in the model. Detailed information about the model can be found in the previous work [30]. The case of single-wavelength amplification is first investigated with the input wavelength set at 1532 nm and the input power set at -20 dBm. It is mentioned that in the simulation the coupling losses of the EDWA chip with external fibers are not considered so the simulated gains and powers correspond to the on-chip values. The simulated output signal spectrum is plotted as red squares in Fig. 5(a), and the input signal power level is shown as dashed blue circles as well. The on-chip gain is about 24 dB by comparing the peak power levels of the input and output signals. This value is consistent with the measured 18 dB off-chip gain considering the 6 dB coupling loss for the fiber-chip-fiber link. Meanwhile, the simulated ASE background is also qualitatively consistent with the experiment. Then the multi-wavelength amplification process is simulated by setting the input signal as the broadband continuous-wave light whose spectrum is shown as blue circles in Fig. 5(b). The amplified output signal spectrum is depicted as red squares in Fig. 5(b), from which the on-chip gain can be retrieved by comparing the red curve with the blue curve. On-chip net gains are all achieved for the input signal wavelength range from 1525 nm to 1580 nm, and the peak gain appears at 1532 nm with the gain variations of 10~18 dB in the C-band (1530 nm to 1565 nm).

The gain spectra evolution with pump powers and waveguide lengths are further simulated for the multi-wavelength amplification. The input signal spectrum is identical to the blue circles in Fig. 5(b). The power of the broadband input signal is set to -20 dBm, which corresponds to the small-signal amplification regime. The pump powers are firstly increased from below 1 mW to 400 mW and the waveguide length is fixed at 8 cm. The simulated gain spectra evolution is shown in Fig. 5(c) with the blue color representing net loss and the red color representing net gain. It can be noticed that the threshold pump power for net gain increases at short wavelengths and the peak gain is concentrated around 1532 nm at high pump powers. A few gain spectra at different pump powers are also shown in Fig. 5(d) for comparison. Next, the gain spectra evolution with increasing waveguide lengths is simulated at the constant pump power of 200 mW and signal power of -20 dBm. In this case, the gain increases with waveguide lengths in the investigated wavelength range, with the prominent gain growth at 1532 nm and the modest gain growths from 1545 nm to 1565 nm. The gain for the wavelength range between 1535 nm and 1542 nm increases slowly with waveguide length, due to the low emission cross sections in this range. Meanwhile, this wavelength range is the lowest gain region in the C-band, greatly limiting the gain flatness of the TFLN-EDWA chip. To filling this gain gap the $Er^{3+}$:LiNbO$_3$ crystal can be co-doped with other elements to homogenize the emission spectra by increasing inhomogeneous broadening.

# Conclusion

To conclude, a compact low-noise TFLN-EDWA chip with 18 dB fiber-to-fiber (off-chip) net gain and 5 dB noise figure is demonstrated. On-chip edge-couplers are employed for high-efficiency coupling with optical fibers. Broadband amplification tests of the EDWA chip reveal the significant gain around 1532 nm due to the inherent spectral properties of $Er^{3+}$:$LiNbO_3$. Theoretical amplifier modeling involving the absorption and emission spectra of $Er^{3+}$ pinpoints the gain-scaling with waveguide lengths for most of the telecom C-band wavelengths. Further improvement on the waveguide configuration and pumping strategy will bring a high-gain EDWA chip for the full telecom C-band. The demonstrated low-noise high-external-gain TFLN-EDWA can be used as compact gain elements in optical communication and integrated lasers with promising high-speed electro-optical tuning abilities.